# Distributed Ride-Matching for Shared Ridehailing Service with Intelligent City Infrastructure

Seyed Mehdi Meshkani
*Laboratory of Innovations in Transportation (LiTrans)*
*Toronto Metropolitan University*
Toronto, Canada
smeshkani@ryerson.ca

Bilal Farooq
*Laboratory of Innovations in Transportation (LiTrans)*
*Toronto Metropolitan University*
Toronto, Canada
bilal.farooq@ryerson.ca

*Abstract*—High computational time is one of the most important operational issues in centralized dynamic shared ridehailing services. To resolve this issue, we propose a distributed ride-matching system that is based on vehicle to infrastructure (V2I) and infrastructure to infrastructure (I2I) communication. The application on downtown Toronto road network demonstrated that the distributed system resulted in a speed-up of 125 times in terms of computational time and showed high scalability. Moreover, the service rate in the proposed system improved by 7% compared to the centralized. However, the centralized system showed 29% and 17% improvement in wait time and detour time, respectively.

## I. INTRODUCTION

Shared mobility is one of the key aspects of Mobility On-Demand (MoD) systems. Shared ridehailing, also known as ridesplitting, where one vehicle can serve multiple passengers simultaneously, is a common and emerging share mobility service that potentially can increase the vehicle occupancy rate. However, to provide a high quality service, numerous operational issues remain to be addressed. Lack of scalability and high computational time is one of the open and under-studied issues in the ride-matching problem associated with the centralized dynamic shared ridehailing system. In such centralized systems, there exists a single control entity that stores the complete data and processes it centrally [1]. To resolve this issue, various decentralized approaches such as decomposition methods and graph partitioning techniques have been used in the literature. These approaches converts the main problem into several sub-problems to reduce the complexity of the ride-matching problem and decrease the computational time [2]–[5]. To further reduce the computational time of the ride-matching problem and make the system highly scalable, in this study, we propose a distributed ride-matching system. Generally, a distributed system consists of multiple servers and storage that are physically separated and located in different locations. They can work together, communicate, and collaborate to achieve the same outcome, creating the illusion of a single, unified system with powerful computing capabilities [6].

This research is funded by the Canada Research Program in Disruptive Transportation Technologies and Services.

The proposed distributed system in this study takes advantage of infrastructure to infrastructure (I2I) and vehicle to infrastructure (V2I) connectivity. The current ride-matching platforms such as Uber, Lyft, and Didi are being performed through a Vehicle to infrastructure (V2I) connectivity for matching, routing, and monitoring purposes. Unlike the current centralized systems, in our proposed distributed system, a network of intelligent intersections ($I^2s$) [7] is utilized as the infrastructure. Each $I^2$ is capable to run a ride-matching algorithm independently based on the data they receive locally. They can also send/receive the data to/from other $I^2s$ as well as the vehicles. To show the efficiency of the proposed distributed system, a Graph-based Many-to-One ride-Matching algorithm (GMOMatch) developed by [8] for shared ridehailing services is applied on it. We then implement the distributed system on the same agent-based traffic microsimulator as used by [8] to estimate different indicators and compare the performance.

To the best of our knowledge, this is the first distributed system for ride-matching problem that takes advantage of I2I and V2I technology. The proposed distributed system can be applied to a variety of MoD services such as ridehailing, shared ridehailing, ridesharing, carsharing, and on-demand transit.

The rest of the paper is organized as follows. In Section II, we briefly review the relevant literature on decentralized/distributed ride-matching systems. Section III proposes the system design and also it provides an analysis on the computational complexity and scalability. Section IV presents the description of the case study, results and discussions. Finally, Section V concludes our findings and provides some directions for future research.

## II. BACKGROUND

In this section, we briefly review the literature on decentralized/distributed ride-matching systems. Reference [3] presented a decentralized auction-based ride-matching algorithm. In their proposed algorithm, participants are divided based on the geographic location. Drivers are matched with passengers whose origins or destinations are in the vicinity of the driver's route. Reference [5] presented a graph partitioning method for one-to-one ridesharing and then they applied the method on many-to-one ridesharing, where one vehicle can serve multiple riders. Although the method was highly efficient for one-to-one ridesharing, no significant improvement was reported in the many-to-one matching problem. Considering the high number of clusters for large size problems was stated as one of the reasons for such a poor performance. Reference [9] proposed a decentralized peer-to-peer ride-matching system to resolve the concerns of privacy and lack of trust among peers in centralized ride-matching systems. They suggested a reputation management protocol through which peers are able to trust each other even when they did not interact before. Reference [1] presented a blockchain-based decentralized service that takes advantage of a time-locked deposit protocol to prevent malicious activities. In their proposed system, a driver should provide proof of being on-time at the pick-up location. Moreover, the elapsed distance of the driver and rider is used as the payment method. Reference [10] proposed a blockchain-based ridehailing system and developed smart contracts to build and deploy functionalities such as create ride, auto-deposit transfer, cancel, and complete ride methods.

Despite some studies deployed partitioning/decomposition methods [5], the structure of their proposed system was still centralized which makes it prone to failures and security attacks. Moreover, they reported unsatisfactory performance when it applied on the many-to-one ride-matching, in which multiple riders can be served by each vehicle. In contrast, in this study we propose a ride-matching system that is structurally distributed and unlike current platforms (e.g. Uber, Lyft) that provide only a centralized V2I communication, it takes advantage of I2I and V2I connectivity. The distributed structure can remarkably decrease the computational time and enables the system to be robust against failure and security issues. Some studies, as reviewed above [1], [10], [11], developed decentralized ride-matching systems utilizing blockchain technology. Despite the benefits of blockchain in developing ride-matching systems, implementation cost and energy consumption can be stated as the key challenges of developing such systems. In our distributed system, however, $I^2s$ can be easily installed and maintained with a reasonable cost and also they are efficient in terms of energy consumption.

## III. METHODOLOGY

In this section, first we design the distributed system and then the GMOMatch algorithm is introduced.

### A. Distributed ride-matching system

A distributed ride-matching system and associated subsystems are developed using the End-to-End dynamic routing for Connected and Automated Vehicles (E2ECAV) platform [7]. E2ECAV is based on a network of intelligent intersections ($I^2$). $I^2s$ are devices installed at intersections and can perform as independent computers. These devices have their own data storage and processors to store and process the data. Moreover, they can communicate with vehicles and other intelligent intersections via Vehicle-to-Infrastructure (V2I) and Infrastructure-to-Infrastructure (I2I) connectivity in order to receive and send information.

In our distributed system (see Algorithm 1), each $I_i^2 \in I^2$ as a local dispatcher is capable of running the ride-matching algorithm independently. In this setup, passenger's request is sent to the nearest $I_i^2$, denoted by $r_{I_i^2}$. Let $N_{I_i^2} \subset N$ be the set of ride requests receiving by a $I_i^2$. Ride requests are placed by passengers over the matching interval, $\Delta$, and then the matching algorithm is run by $I_i^2$ at each matching time. Each $I_i^2$ starts the search process to find available vehicles $v_{I_i^2}$ nearby. $V_{I_i^2} \subset V$ represents the set of available vehicles recognized by $I_i^2$. The procedure of searching can be locally on the $I_i^2$'s inbound links—we name it search level zero, denoted by $S_0$ ($I_1^2$ in Fig. 1a). $I_i^2$ can also communicate with its neighbours through I2I connectivity to enlarge the search space. Two $I_i^2$s are called neighbours when they are directly connected to each other by a link. For instance, in Fig. 1b, the $I_1^2$ has four neighbours, including $I_2^2, I_3^2, I_4^2, I_5^2$. Each neighbour starts searching locally to find available vehicles and send their information to the $I_1^2$. Depending on the level at which an $I_i^2$ expands its search space, three more search levels are defined ($S_1, S_2, S_3$). It is worth mentioning that the search level remains the same over each simulation scenario. For example, with search level $S_0$ in a scenario, every time the $I_1^2$ runs the matching algorithm at each matching time, it searches locally to find available vehicles. However, when the search level is set to $S_1$, at each matching time, $I_i^2$ searches locally and also communicates with all of its neighbors, including $I_2^2, I_3^2, I_4^2, I_5^2$ to find available vehicles. Vehicles are connected to the $I_1^2$s (V2I) either through an application on the driver's smart phone or an electronic device installed on the vehicle.

Scalablility is one of the main features of the proposed distributed system. Each $I^2$ is an independent computer and it is easy to add or remove $I^2$ from the system. Moreover, every $I^2$ in the architecture is allowed to enter or exit at any time. As a result, distributed systems have higher levels of reliability in terms of fault-tolerant. Furthermore, the proposed distributed system allows multiple $I^2s$ to work on the same process, improving the performance of such systems. Through this load balancing, operations will be faster and more cost-efficient.

---

**Algorithm 1**

**Input:** Set of passengers $N_{I_i^2}$ associated with $I_i^2$, spatiotemporal information of passengers
**Output:** Passengers-vehicles matching
$K \leftarrow |I^2|$
**while** $i \leq K$ **do**
  $N_{I_i^2} \leftarrow r_{I_i^2}$
  $I_i^2$ starts searching for available vehicles:
  $V_{I_i^2} \leftarrow v_{I_i^2}$
  $I_i^2$ runs the ride-matching algorithm
**end while**

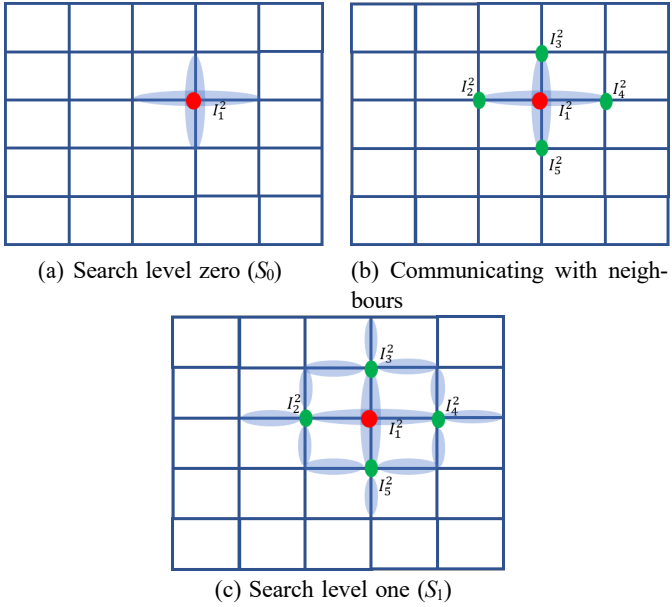

(a) Search level zero ($S_0$)  (b) Communicating with neighbours

(c) Search level one ($S_1$)

Fig. 1: Searching process in the distributed system

### B. GMOMatch algorithm

The GMOMatch ride-matching algorithm was developed by [8] for shared ridehailing services where one vehicle can serve multiple passengers simultaneously. The proposed algorithm, which is a graph-based heuristic algorithm, is iterative and includes two steps. It starts with one-to-one matching in step 1 and is followed by matching of vehicles to combine passengers with similar itineraries in step 2. In order to do one-to-one matching in step 1, a bipartite graph is created and then solved using Hungarian algorithm. The assigned vehicles from step 1 form a directed general graph, called vehicle graph, in step 2. In the obtained vehicle graph, to match the vehicles and combine passengers with similar itineraries, a maximum weight matching problem needs to be solved. Edmonds' algorithm was utilized to solve this problem. The objective function of both bipartite graph and vehicle graph is minimizing vehicles travel time. The output of the algorithm is matching passengers with vehicles and determining the sequence of pick up/drop off of the passengers for each vehicle. According to [8], the GMOMatch provided a high-quality service. However, to reduce the computational time, the algorithm is implemented on the proposed distributed system that enables each $I^2 \in I^2$ to run the GMOMatch independently.

### C. Complexity and scalability analysis

In the original GMOMatch, according to [8], the complexity of creating bipartite graph in step 1 of the algorithm is $O(mks^2)$, where $m$ is the number of requests, $k$ is the maximum number of feasible vehicles, and $s$ is the number of feasible spots in the insertion method. The complexity of the Hungarian algorithm to solve bipartite graph is $O(N^3)$ where $N$ is the size of cost matrix. In Step 2 of the algorithm, the complexity of the entire vehicle graph is $O(n'k's'^2)$ where $n'$ is the number of assigned vehicles, $k'$ is the maximum feasible vehicles and $s'$ is the number of feasible spots in the insertion method associated with step 2. The complexity of Edmonds' algorithm in a general graph is $O(V^2E)$ where $V$ represents number of vertices and $E$ is the number of edges. Thus, the computational complexity of the entire algorithm is $O(max \{mks^2, N^3, n'k's'^2, V^2E \})$.

Given the fact that in the proposed distributed system, each intelligent intersection is a local dispatcher that performs independently, the computational complexities of the distributed version of GMOMatch is $O(max\{\frac{mks^2}{\zeta}, (\frac{N}{\zeta})^3, \frac{n'k's'^2}{\zeta}, \frac{V^2E}{\zeta}\})$, where $\zeta$ is the number of intelligent intersections that receives ride request.

Scaling in the proposed distributed system can be vertically or horizontally [12]. In horizontal scaling, more $I^2$s are added to (or removed $I^2$s from) the distributed system while vertical scaling refers to adding resources to (or removing resources from) a single $I^2$, such as adding CPUs, memory, or storage. If $\beta$ is the portion of a calculation that is sequential, and 1-$\beta$ is the portion that can be parallelized, the maximum speedup that can be achieved by using K processors is estimated according to Amdahl's Law [13] (Eq. 1):

$$\frac{1}{\beta + \frac{1-\beta}{K}} \quad (1)$$

For instance, given 75% of a program can be sped up if parallelized, using 4 processors gives:

$$\frac{1}{0.25 + \frac{1-0.25}{4}} = 2.285 \quad (2)$$

and using 8 processors gives:

$$\frac{1}{0.25 + \frac{1-0.25}{8}} = 2.909 \quad (3)$$

By doubling the processing power, the process was only sped up by 27%. This analysis indicates that vertical scaling and adding more processors to $I^2$s is not necessarily the optimal approach. In Section IV, we will show the high efficiency of horizontal scaling by adding more $I^2$s.

## IV. CASE STUDY AND RESULTS

The road network of Downtown Toronto is used as the study area. Fig. 2 shows a partial as well as a full road network of Downtown Toronto. The partial network consists of 76 nodes/intersections, 223 links with the size of 0.70km x 2.61km, while the full network contains 268 nodes/intersections, 839 links with the size of 3.14km x 3.31km.

The dynamic demand loading period was 7:45 am–8:00 am (15 minutes) in the morning peak period. To synthesize the data, 2011 Transportation Tomorrow Survey (TTS) travel data for city of Toronto was used while the 2018 growth factor was applied. The total number of trips in the loading period for the full and partial network was 5,487 and 3,477 trips, respectively. To select the demand of shared vehicles, a specific percent of the total demand were randomly extracted while the rest of the demand was assumed to travel by their own single-occupancy private vehicles. Despite the fact that the dynamic demand loading period was 15 minutes, the simulation time lasted until all passengers either arrived at their destination or left the system. All simulations in this study were run using two similar computers with Core i7-8700 CPU, 3.20 GHz Intel with a 64-bit version of the Windows 10 operating system with 16.0 GB RAM.

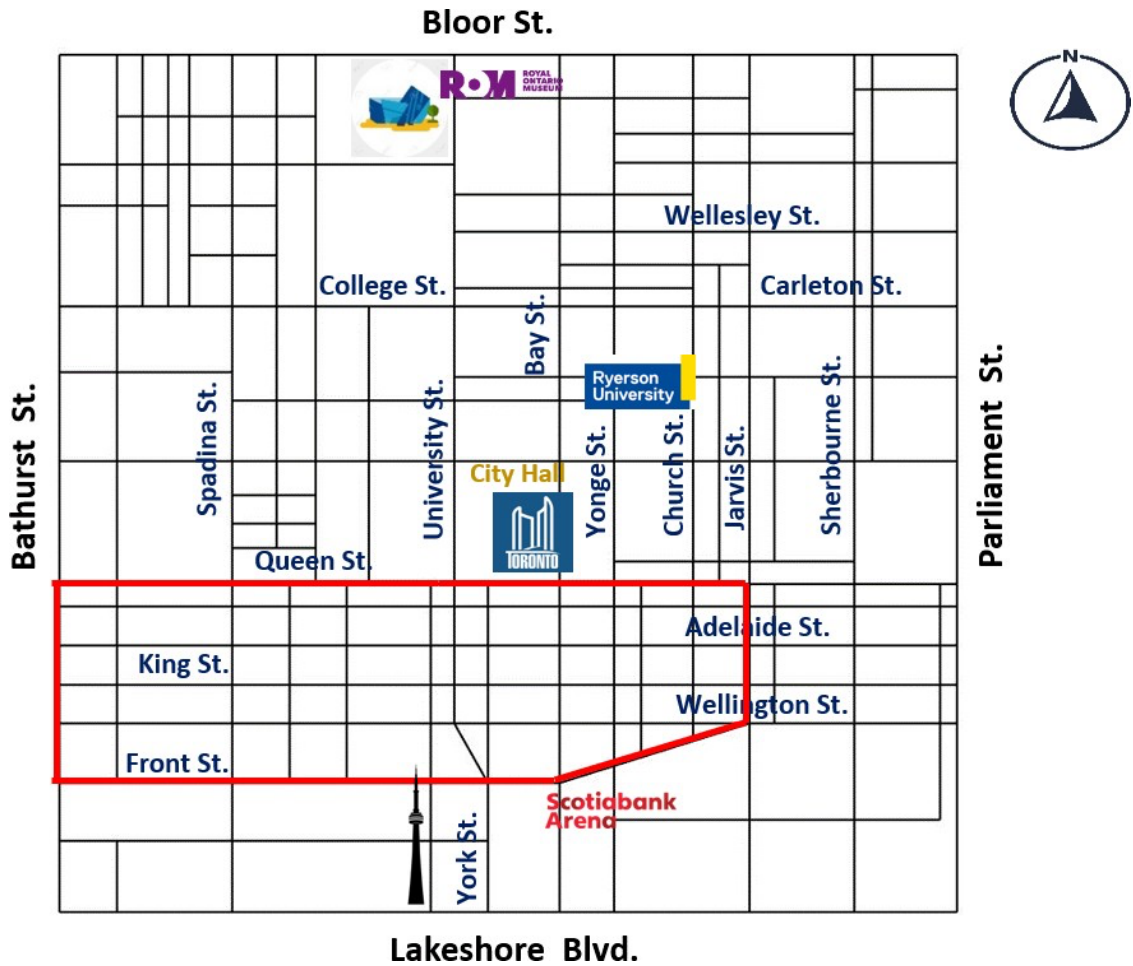

Fig. 2: Downtown Toronto street network

For the sake of comparison, the proposed distributed version of GMOMatch was implemented in the same agent-based traffic microsimulator used by the original GMOMatch [8], which is centralized. To evaluate the performance of the proposed system, five scenarios were created. These scenarios include four search levels that were compared with the centralized GMOMatch as the base case.

Fig. 3 showcases the results of this comparison over different indicators, including service rate (SR), computational time (CT), wait time (WT), detour time (DT), traffic travel time (TTT), and vehicle kilometer traveled (VKT). The shared vehicles demand was considered to be 25% of the total demand, which came out to be 1,097 trips, and three fleet sizes (SVs) of 230, 250, and 270 were used. The matching interval was $\Delta = 30$ sec and the capacity of vehicles (*cap*) is four.

Fig. 3a shows the SR for different scenarios. As can be seen, the distributed system for different search levels over various fleet size show similar or higher SR compared to the centralized version. The SR for search level 03 with $SV = 270$ is 95% while it is 89% for the centralized. Fig. 3b reveals the CT. According to the figure, there is a significant difference between centralized and various search levels of the distributed system such that for $S_3$ with $SV = 270$, the CT is 1.22 sec while for the centralized it is 152.14 sec, which is a speed-up of 125 times. The reason for such a large difference is that in the centralized, all of the calculations are computed by a single dispatcher, while in the distributed system these calculations are divided among the total number of intelligent intersections

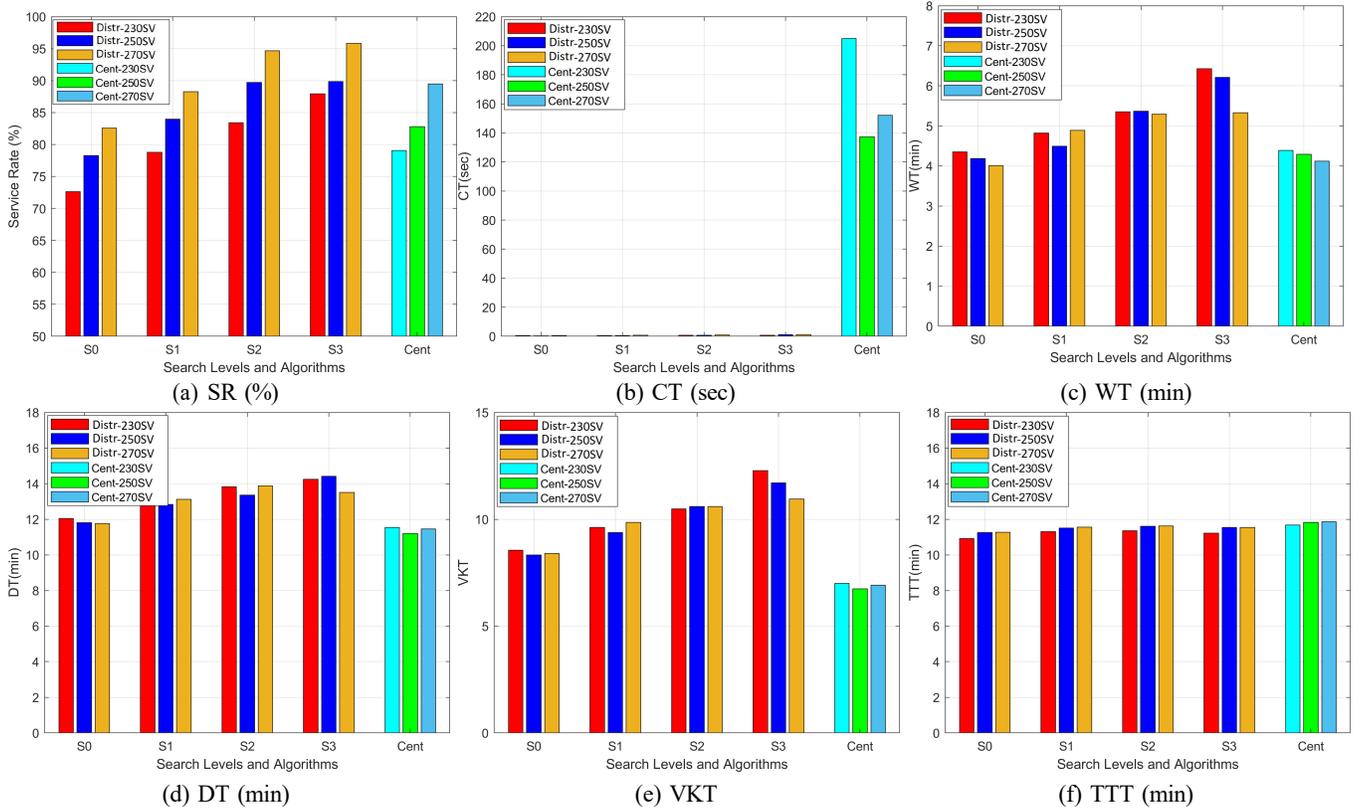

(a) SR (%)  (b) CT (sec)  (c) WT (min)
(d) DT (min)  (e) VKT  (f) TTT (min)

Fig. 3: distributed vs centralized: demand=25%, cap=4, $\Delta = 30 sec$

that receive ride request.

Although the distributed system outperforms the centralized in terms of SR and CT, the centralized shows a better performance for WT, DT, and VKT. Fig. 3c and Fig. 3d demonstrates the WT and DT for different scenarios. The centralized GMOMatch showed 29% and 17% improvement in the WT and DT compared to the S3 of distributed with SV=270. Moreover, according to Fig. 3e, VKT for centralized GMOMatch is 60% lower than the S3 of distributed with SV=270.

Fig. 3f represents traffic travel time in the network. As can be seen, there is a slight difference between centralized and distributed version of GMOMatch meaning that they almost have similar impact on traffic congestion.

To show the scalability of the proposed distributed system, which is horizontally, we applied it on the partial downtown road network in Fig. 2 and compared the CT with the full network. Fig. 4 showcases the performance of distributed system in terms of the CT and the SR for both partial and full networks. The demand for shared vehicles was 20% of each network's total demand, which was 695 and 1,097 trips for the partial and full network, respectively. The number of shared vehicles was 75 and 150 for each network. As expected, by enlarging the search space, the SR is increased such that for search level 03, this indicator raised to 90.67% and 91.59% for the partial and full network, respectively. However, by scaling up the network, no significant difference was observed in the CT over various search levels.

Overall, the obtained results indicate that the proposed distributed system is able to effectively figure out the issue of high computational time and lack of scalability of centralized version of GMOMatch. It also provides a high SR that demonstrate how successful the system is in dispatching travellers. However, the centralized system due to having a global view of the network yields a higher service quality compared to the distributed system.

## V. CONCLUSION

In this study, we proposed a novel distributed system to tackle the issue of high computational time in the ride-matching problem associated with the centralized shared ride-hailing service. The proposed distributed system performs based on infrastructure to infrastructure (I2I) and vehicle to infrastructure (V2I) connectivity. The infrastructure is able to run a ride-matching algorithm based on the data received locally. To evaluate the performance, the GMOMatch algorithm developed by [8] was implemented on the distributed system and compared with the original centralized version. To our knowledge, this is the first distributed system for ride-matching using I2I and V2I technologies.

An analysis on computational complexity of both systems showed that utilizing intelligent intersections as local dispatchers can remarkably reduce the complexity of the original

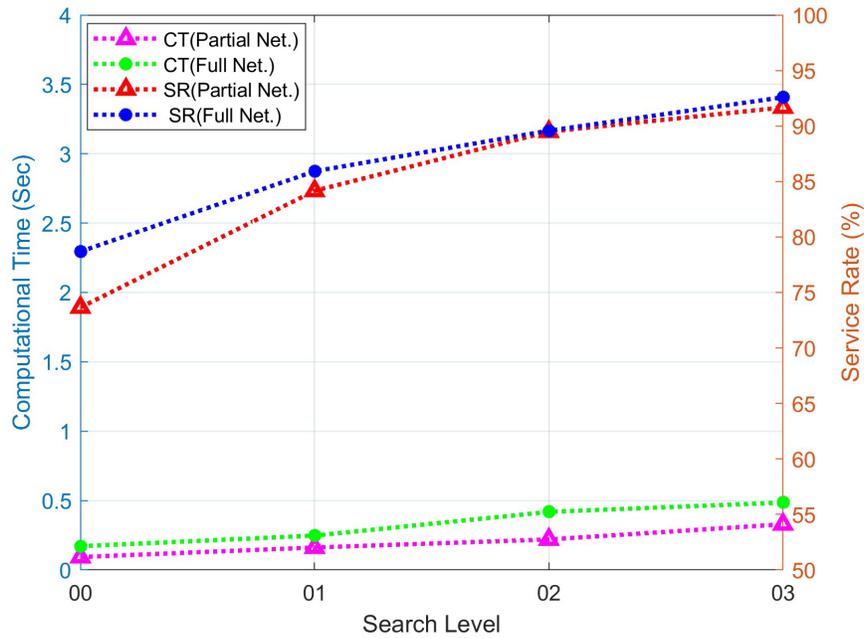

Fig. 4: Scalability of the system. partial network: SV demand=695 trips, SV=75, cap=4, $\Delta = 60sec$. full network: SV demand=1097 trips, SV=150, cap=4, $\Delta = 60sec$

centralized GMOMatch. Moreover, the results on Downtown Toronto revealed that deploying V2I and I2I connectivity in the proposed distributed system brings a speed-up of 125 times. Moreover, the service rate was 95%, which was 6.00% higher than centralized system. However, for some other indicators, including wait time, detour time, and vehicle kilometer travelled, the centralized system provided a better performance by 29%, 17%, and 60%, respectively.

The proposed distributed system could be useful if the shared ridehailing service suddenly becomes overloaded and receives many ride requests within a short period of time. In downtown Toronto, teams such as the Maple Leafs (ice hockey), Blue Jays (baseball) and Toronto Raptors (basketball) have stadiums, and after games, there is a high demand for getting a ride in the surrounding intersection(s). As in distributed systems, different components can collaborate and process the same task, intelligent intersections can work together in such a scenario to dispatch ride requests efficiently and keep travellers' wait times at a reasonable level. It is worth mentioning that the proposed distributed system is applicable to a variety of MoD services, including ridesharing, carsharing, and on-demand transit.

For future research, to enhance the service quality of the proposed distributed system, more advanced methods such as game theory can be used to enhance the cooperation among intelligent intersections.

## References

[1] M. Baza, M. Nabil, M. Ismail, M. Mahmoud, E. Serpedin, and M. A. Rahman, "Blockchain-based charging coordination mechanism for smart grid energy storage units," in *2019 IEEE International Conference on Blockchain (Blockchain)*. IEEE, 2019, pp. 504–509.
[2] N. Agatz, A. Erera, M. Savelsbergh, and X. Wang, "Optimization for dynamic ride-sharing: A review," *European Journal of Operational Research*, vol. 223, no. 2, pp. 295–303, 2012.
[3] M. Nourinejad and M. J. Roorda, "Agent based model for dynamic ridesharing," *Transportation Research Part C: Emerging Technologies*, vol. 64, pp. 117–132, 2016.
[4] A. Najmi, D. Rey, and T. H. Rashidi, "Novel dynamic formulations for real-time ride-sharing systems," *Transportation research part E: logistics and transportation review*, vol. 108, pp. 122–140, 2017.
[5] A. Tafreshian and N. Masoud, "Trip-based graph partitioning in dynamic ridesharing," *Transportation Research Part C: Emerging Technologies*, vol. 114, pp. 532–553, 2020.
[6] G. Coulouris, J. Dollimore, and T. Kindberg, "Distributed systems: Concepts and design edition 3," 2001.
[7] B. Farooq and S. Djavadian, "Distributed traffic management system with dynamic end-to-end routing," U.S., 2019, u.S. provisional patent 62/865,725.
[8] S. M. Meshkani and B. Farooq, "A generalized ride-matching approach for sustainable shared mobility," *Sustainable Cities and Society*, vol. 76, p. 103383, 2022.
[9] D. Sa´nchez, S. Mart´ınez, and J. Domingo-Ferrer, "Co-utile p2p ridesharing via decentralization and reputation management," *Transportation Research Part C: Emerging Technologies*, vol. 73, pp. 147–166, 2016.
[10] S. Kudva, R. Norderhaug, S. Badsha, S. Sengupta, and A. Kayes, "Pebers: Practical ethereum blockchain based efficient ride hailing service," in *2020 IEEE International Conference on Informatics, IoT, and Enabling Technologies (ICIoT)*. IEEE, 2020, pp. 422–428.
[11] H. Yu, V. Raychoudhury, and S. Silwal, "Dynamic taxi ride sharing using localized communication," in *Proceedings of the 21st International Conference on Distributed Computing and Networking*, 2020, pp. 1–10.
[12] M. Michael, J. E. Moreira, D. Shiloach, and R. W. Wisniewski, "Scale-up x scale-out: A case study using nutch/lucene," in *2007 IEEE International Parallel and Distributed Processing Symposium*. IEEE, 2007, pp. 1–8.
[13] R. E. Bryant, O. David Richard, and O. David Richard, *Computer systems: a programmer's perspective*. Prentice Hall Upper Saddle River, 2003, vol. 2.